\documentclass[12pt]{article}

\usepackage{amssymb}
\usepackage{amsmath}
\usepackage{amscd}
\usepackage{latexsym}

\usepackage{url}

\usepackage{cite}

\usepackage{graphicx}
\usepackage{subfigure}

\newcommand{\be}{\begin{equation}}
\newcommand{\ee}{\end{equation}}

\newcommand{\dlt}{\delta}
\newcommand{\prt}{\partial}
\newcommand{\br}{{\bf r}}

\newcommand{\ep}{\varepsilon}
\newcommand{\al}{\alpha}

\newcommand{\om}{\omega}
\newcommand{\Om}{\Omega}

\newcommand{\rgl}{\rangle}
\newcommand{\lgl}{\langle}

\begin{document}

\begin{center}

{\Large{\bf Realization of inverse Kibble-Zurek scenario with trapped Bose gases} \\ [5mm]

V.I. Yukalov$^{1,2}$, A.N. Novikov$^{1,2}$, and  V.S. Bagnato$^1$,} \\ [3mm]

{\it
$^1$Instituto de Fisica de S\~{a}o Calros, Universidade de S\~{a}o Paulo, \\
CP 369, 13560-970 S\~{a}o Carlos, S\~{a}o Paulo, Brazil \\ [3mm]

$^2$Bogolubov Laboratory of Theoretical Physics, \\
Joint Institute for Nuclear Research, Dubna 141980, Russia} \\ [3mm]

\end{center}

\vskip 2cm

\begin{abstract}

We show that there exists the {\it inverse Kibble-Zurek scenario}, when we 
start with an equilibrium system with broken symmetry and, by imposing 
perturbations, transform it to a strongly nonequilibrium symmetric state 
through the sequence of states with spontaneously arising topological defects. 
We demonstrate the inverse Kibble-Zurek scenario both experimentally, by 
perturbing the Bose-Einstein condensate of trapped $^{87}$Rb atoms, and also 
by accomplishing numerical simulations for the same setup as in the experiment, 
the experimental and numerical results being in good agreement with each other.

\end{abstract}

\vskip 2mm

{\it Keywords}: Nonequilibrium Bose gas; Quantum vortices; Vortex turbulence;
                Grain turbulence, Wave turbulence; Inverse Kibble-Zurek scenario 

\vskip 1cm

{\parindent =0pt
{\bf Corresponding author}: V.I. Yukalov, \\ [2mm]
{\bf Address}: Bogolubov Laboratory of Theoretical Physics, \\ 
Joint Institute for Nuclear Research, Dubna 141980, Russia. \\ [2mm]
{\bf Phones}: 7(496)216-3947; 7(496)213-3824 \\
{\bf E-mail}: yukalov@theor.jinr.ru 
}   

\newpage

\section{Introduction}

The Kibble-Zurek mechanism \cite{Kibble_1,Kibble_2,Zurek_3,Zurek_4,Zurek_5} 
characterizes the spontaneous formation of defects in the process of system
equilibration from an initial strongly nonequilibrium symmetric state to an
equilibrium state with broken symmetry. This is a very general mechanism 
illustrated by numerous examples ranging from various condensed-matter 
systems to cosmological objects \cite{Kibble_2,Zurek_4,Zurek_5}. It has been 
experimentally observed, in ion crystals \cite{Pyka_6,Ulm_7}, superconducting 
films \cite{Carmi_8}, superfluid $^3$He and superfluid $^4$He, modeling 
cosmological string formation \cite{Bauerle_9,Ruutu_10,Hendry_11}, in trapped 
atoms exhibiting Bose-Einstein condensation \cite{Weiler_12,Lamporesi_13},
and in a homogeneous Bose gas \cite{Navon_46}.       

The case of nonequilibrium Bose-Einstein condensation in dilute Bose gases 
has attracted special attention since the work by Levich and Yakhot 
\cite{Levich_14} followed by many other publications studying, both 
theoretically \cite{Stoof_15,Kagan_16,Zakharov_17} and in computer simulations 
\cite{Snoke_18,Semikoz_19,Berloff_20}, the process of equilibration of an 
uncondensed system to its Bose-condensed state. The general understanding of 
the process of this equilibration, passing through the states with topological 
defects, is as follows. 

One starts with a strongly nonequilibrium state possessing no Bose-Einstein 
condensate, which is termed wave turbulence or weak turbulence. In the process 
of equilibration, the Kibble-Zurek mechanism is exhibited, first, by the 
appearance of local superfluid regions of condensate, with the characteristic 
size of coherence length, being yet not correlated with each other. These 
regions, called by Kibble \cite{Kibble_1,Kibble_2} protodomains or cells, in 
the case of Bose condensation, are the germs, or grains, or droplets of 
Bose-Einstein condensed atoms inside uncondensed surrounding. Such a granular 
state represents the stage of strong turbulence, which can be called grain 
turbulence in order to stress the physical origin of the objects forming the 
state. This is similar to naming weak turbulence as wave turbulence for 
emphasizing that it is formed by small-amplitude waves.

In the next stage of equilibration, the number of grains, or cells, grows so 
that they start fusing with each other forming numerous vortices 
\cite{Kibble_1,Kibble_2,Weiler_12}. When no anisotropy is imposed on the system,
the created vortices form a random tangle corresponding to vortex turbulence
\cite{Feynman_21}. 

In the process of the following equilibration, the number of vortices 
diminishes, and the system passes to a vortex state with several vortices, 
whose number continues decaying, so that finally the system relaxes to its 
equilibrium Bose-condensed state without vortices. 

In the present paper, we demonstrate, both experimentally and by computer 
simulations, that it is feasible to realize the {\it inverse Kibble-Zurek 
scenario}, starting with an equilibrium Bose-Einstein condensate, exciting 
which, we go through the sequence of states with spontaneously arising 
topological defects, as in the equilibration process, but in the inverse 
order. Perturbing the system, we subsequently transfer it into the states 
of weak nonequilibrium, with no vortices, then to a vortex state, with 
several vortices, then to the states of vortex turbulence, grain turbulence, 
and, finally, to the state of wave turbulence, where the condensate is 
completely destroyed. 

This is demonstrated by considering a trapped Bose-Einstein condensate, 
perturbing which we realize the whole sequence of states whose order is 
opposite to that arising in the process of equilibration of an uncondensed 
system to its Bose condensed state. All intermediate states with topological 
defects are recovered, although in the opposite to the Kibble-Zurek mechanism 
order.

Although we call the process of transferring a Bose gas from its equilibrium 
state to a strongly nonequilibrium state {\it inverse Kibble-Zurek scenario}, 
this does not necessarily imply that all particular details of this mechanism 
have to be exactly the same, but in the opposite order. Moreover, there is an
important difference. One often considers the direct Kibble-Zurek mechanism 
caused by quenching the temperature of an external bath. However, in our case, 
there is no external bath fixing or varying temperature. We realize the transfer 
from an equilibrium state to highly nonequilibrium states by strong time-dependent
perturbations of a {\it finite closed system} that is not connected to any bath. 
So, our system is strongly nonequilibrium. This principally distinguishes it 
from the situation of a stationary system, where a kind of vortices can be 
thermally excited in the vicinity of the Bose-condensation temperature 
\cite{Williams_47,Antunes_48,Nemirovskii_49,Nemirovskii_50}. These thermally
excited vortices in a stationary system cannot form turbulence, while in our 
case of a strongly nonequilibrium system, vortex turbulence does happen. 

Naming the process, we describe, as the inverse Kibble-Zurek scenario, we aim 
at emphasizing that the created nonequilibrium states with topological defects 
occur in the order opposite to the direct Kibble-Zurek mechanism. However, this 
does not mean the exact correspondence to all particularities of this mechanism. 
More precisely, the process, we realize, is inverse with respect to the so-called 
nonequilibrium Bose-Einstein condensation, because of which it could be termed 
the {\it inverse Bose-Einstein condensation}. However, we wish to stress the 
general possibility of reverting, not only Bose condensation, by also other 
different nonequilibrium phase transitions, realizing 
{\it inverse nonequilibrium transitions}.

\section{Perturbation of Bose-Einstein condensate}

We employ the same experimental setup as in the experiments studying quantum 
turbulence of trapped $^{87}$Rb atoms \cite{Henn_22,Shiozaki_23,Seman_24}. 
The gas of $^{87}$Rb atoms is cooled down to the state where almost all atoms, 
up to $70 \%$ are condensed, with the number of atoms in Bose-Einstein 
condensate being $N \approx 2 \times 10^5$. The harmonic trap has cylindrical 
shape, with characteristic frequencies $\omega_\perp = 2 \pi \times 210$ Hz and
$\omega_z = 2 \pi \times 23$ Hz. Then the aspect ratio
\be
\label{1}
 \al \equiv \frac{\om_z}{\om_\perp} = 
\left ( \frac{l_\perp}{l_z} \right )^2 \;  ,
\ee
where $l_{\perp}$ and $l_z$ are the transverse and axial oscillator lengths,
is $\alpha = 0.11$.  

After forming equilibrium Bose-Einstein condensate, the atomic cloud is subject 
to an oscillatory trap modulation that shakes the condensate without imposing 
a rotation axis, but preserving the cloud isotropy \cite{Yukalov_25}. As in our 
previous papers \cite{Henn_22,Shiozaki_23,Seman_24}, we use the perturbing 
potential 
\be
\label{2}
 V(\br,t) = \frac{m}{2} \; \Om_x^2(t) (x' - x_0')^2 +
\frac{m}{2} \; \Om_y^2(t) ( y' - y_0')^2 + \frac{m}{2} \; 
\Om_z^2(t) (z' - z_0')^2 \; ,
\ee
in which $m$ is atomic mass, $x', y', z', x'_0, y'_0, z'_0$ are the variables 
shifted and tilted with respect to the real spatial variables $x, y, z$, and 
$\Omega_\alpha(t) = A_\alpha \omega_\alpha [1 - cos(\omega t)]$ are alternating 
frequencies oscillating with an effective amplitude $A_\alpha$ and frequency 
$\omega = 2 \pi \times 200$ Hz. This perturbing potential distorts the atomic 
cloud with a combination of deformations, rotations, and displacements 
\cite{Henn_22,Shiozaki_23,Seman_24}. The modulation amplitude and time can be 
varied in a wide range, thus, varying the energy injected into the system. The
injected energy per atom can be evaluated \cite{Birman_33,Bagnato_41} as  
\be
\label{3}
 E_{inj} = \frac{1}{N} \; \int_0^t \left | \left \lgl 
\frac{\prt\hat H}{\prt t} \right \rgl \right | \; dt \;  ,
\ee
expressed through the energy Hamiltonian 
\be
\label{4}
  \hat H = \int \eta(\br,t) \left ( -\; \frac{\nabla^2}{2m} + U 
\right ) \eta(\br,t) \; d\br \; + \; 
\frac{1}{2} \; \Phi_0 \int | \eta(\br,t)|^4 \; d\br \;.
\ee
Here $\eta = \eta(\bf{r},t)$ is the condensate wave function, $U = U(\bf{r},t)$
is the total external potential
\be
\label{5}
 U(\br,t) = U(\br) + V(\br,t) \;  ,
\ee
including a trap potential $U(\bf{r})$ and the modulation potential $V(\bf{r},t)$.
The atomic interaction strength is
\be
\label{6}
 \Phi_0 = 4\pi \; \frac{a_s}{m} \; ,
\ee
with $a_s$ being scattering length and $m$, atomic mass. We use the system of 
units, where the Planck constant $\hbar$ is set to one. 

The atomic cloud at initial time is assumed to be almost completely condensed,
so that the condensate function is normalized to the total number of atoms
\be
\label{7}
 N = \int | \eta(\br,0) |^2 \; d\br \; .
\ee
The equation of motion for the condensate wave function is
\be
\label{8}
 i\; \frac{\prt \eta}{\prt t} = \left ( -\; \frac{\nabla^2}{2m} + U +
\Phi_0 |\eta|^2 \right ) \eta \;  .
\ee

Increasing the amount of the injected energy $E_{inj}$ disturbs the ground-state
condensate and generates a sequence of nonequilibrium states. Although the 
created nonequilibrium states can be compared to some equilibrium states 
\cite{Yukalov_26}, strictly speaking, they are quite different. As is 
emphasized in Introduction, in equilibrium systems, there can be no turbulence,
while in our finite closed system turbulent motion does appear.  

Observations were done in time-of-flight experiments, after switching off the 
trapping potential. In the typical time-of-flight procedure, the chaotization 
\cite{Brezinova_27} and nonadiabatic \cite{Yukalov_28} effects do not occur
and the expanding atomic cloud preserves the features of the state inside the 
trap \cite{Henn_22,Shiozaki_23,Seman_24}.    

We also accomplished numerical simulations for the whole process of producing 
strongly nonequilibrium states, with the setup and all parameters being exactly 
the same as in the experiment with $^{87}$Rb. This was done by solving the 
nonlinear Schr\"{o}dinger equation (8) by the methods of static \cite{Blum_29} 
and dynamic \cite{DeVries_30} simulations. The dissipation is taken into 
account by introducing the phenomenological factor $\gamma$ 
\cite{Tsubota_31,Nemirovskii_32}. This dissipation slightly diminishes the 
number of atom in time, although not more than $10 \%$. Increasing the amount 
of the injected energy, we have reproduced the whole sequence of nonequilibrium 
states: weak nonequilibrium, vortex state, vortex turbulence, grain turbulence, 
and wave turbulence. All numerical results are in good agreement with 
experimental data. 
 
By means of the alternating perturbation potential, modulating the trap, we 
inject into the trap the energy per atom $E_{inj}$. By increasing the amount 
of the injected energy, we observe the following sequence of nonequilibrium 
states.

\subsection{Weak nonequilibrium} 

Under weak perturbations, only elementary collective excitations arise, with 
the cloud bending and center-of-mass dipole oscillations, but without 
topological defects. Such elementary excitations are typical of collective 
excitations arising in weakly perturbed finite quantum systems \cite{Birman_33}. 
These excitations in Bose gases have been widely studied \cite{Pethick_34}.
So that we do not need to spend time for their discussion. 

In this regime, the average kinetic energy is yet much lower than the interaction 
energy. The effective interaction parameter is defined as 
\be
\label{9}
 g \equiv 4\pi N \; \frac{a_s}{l_\perp} \;  .
\ee
This parameter is large because of the large number of atoms $N$, allowing for 
the use of the Thomas-Fermi approximation, where the effective atomic cloud 
radius and length are
\be
\label{10}
  R = \left ( \frac{15}{4\pi} \; \al g \right )^{1/5} l_\perp =
1.036 \; l_\perp (\al g)^{1/5} \; , \qquad
\al L = 2R = 2.072 \; l_\perp (\al g)^{1/5} \;  .
\ee
We shall use these expressions in our following estimates.

\subsection{Vortex state}. 

The regime of weak perturbation lasts till the injected energy reaches the 
value $2E_{vor}$ required for generating a vortex-antivortex pair. Under the
employed perturbation potential, imposing no rotation axis, vortices usually 
arise in pairs or larger groups, such that the total rotation moment be close 
to zero. 

The energy of a vortex, per atom, in the Thomas-Fermi approximation, 
\cite{Pethick_34}, reads as
\be
\label{11}
 E_{vor} = \frac{5}{2m R^2} \; \ln \left ( 3.364 \; R \sqrt{\rho a_s} \right ) \; ,
\ee
where $\rho$ is atomic density. Depending on the value of the density, the 
number under the logarithm can be slightly different, which, however, does not 
change much the vortex energy. For example, the average density can be defined
as $N/\pi R^2 L$. But for a vortex along the axis of the trap, the central 
density is
\be
\label{12}
 \rho = \frac{15 N}{4\pi R^2 L} = 
0.537 \; \frac{\al N}{l_\perp^3 (\al g)^{3/5} } \;  .
\ee  
When there are several vortices, the related densities $\rho$ are different, 
because the density of the atomic cloud in a trap varies in space 
\cite{Pethick_34,Courteille_35,Yukalov_36}.
However, this variation does not essentially change the related vortex energy, 
since the density is inside the logarithm. Therefore, to estimate the vortex 
energy, we can accept density (12), which gives    
\be
\label{13}
E_{vor} = 0.93 \; \frac{\om_\perp}{(\al g)^{2/5} } \; \ln ( 0.44 \al g ) \; .   
\ee

In the present setup, $\alpha = 0.11$ and $g = 1.96 \times 10^4$. The transverse 
trap energy $E_\perp \equiv \hbar \omega_\perp = 0.869 \times 10^{-12}$ eV, 
which gives $E_{vor} = 0.257 \times 10^{-12}$ eV. In units of the 
transverse trap energy, $E_{vor} = 0.296 E_\perp$. Hence, 
$2E_{vor} = 0.591 E_\perp$. 

After the injected energy surpasses $2E_{vor}$, the acting perturbation 
starts generating vortices. Since no single rotation axis is imposed onto 
the system, the creation of vortices and anti-vortices is equally likely. 
The typical situation is that vortices and anti-vortices appear together 
in groups \cite{Henn_22,Shiozaki_23,Seman_24}. The vortex-antivortex 
pairs do not form bound dipoles, being not correlated with each other, 
except having opposite vorticities. As we have checked by numerical 
simulations, practically all vortices have vorticity one or minus one, 
so that the total momentum is close to zero. The vorticity was calculated
by the method described in Ref. \cite{Foster_42}.

\subsection{Vortex turbulence}

Pumping into the trap more energy creates more and more vortices and 
anti-vortices. The used perturbing potential does not impose a rotation axis, 
because of which the increasing number of vortices does not lead to the 
creation of a vortex lattice, but results in the formation of a random vortex 
tangle representing, according to Feynman \cite{Feynman_21}, quantum vortex 
turbulence \cite{Tsubota_31,Nemirovskii_32}. In this regime, the vorticity 
vectors are randomly oriented \cite{Yukalov_43}. This regime develops, when 
the number of vortices $N_c$ is such that the mean distance between their 
surfaces becomes twice the coherence length $\xi \sim \hbar /m c$, where 
$c \sim (\hbar/m) \sqrt{4 \pi \rho a_s}$ is sound velocity. This gives
\be
\label{14}
 N_c = \left ( \frac{r_\perp}{4\xi} \right )^2 \;  ,
\ee
where $r_\perp$ is the atomic cloud transverse radius. For the treated case 
of $^{87}$Rb atoms, $\xi \sim 2 \times 10^{-5}$ cm and 
$r_\perp \approx 4 \times 10^{-4}$ cm. Then this estimate gives 
$N_c \approx 25$, which is in agreement with both numerical simulations and 
experiment. 

The number of vortices as a function of time, that is, as a function of the
injected energy, under a fixed perturbation amplitude, is shown in Fig. 1,
found in numerical simulations. As is seen, when the number of vortices 
reaches about $25$ there occurs a change in the behavior of the function $N(t)$, 
which is connected with the arising vortex turbulence. The following changes 
in the function $N(t)$ will be discussed below.  

In the time-of-flight observation, the turbulent atomic cloud expands 
isotropically \cite{Henn_22,Shiozaki_23,Seman_24}. This suggests that the 
observed random vortex turbulence is a kind of Vinen turbulence
\cite{Tsubota_31,Nemirovskii_32}. Although a more precise classification of 
the turbulence type requires the study of spectra, which is yet in progress.  

The energy required for creating the random vortex turbulence can be estimated 
as $E_{tur} \sim N_c E_{vor} \sim 0.643 \times 10^{-11}$ eV, or in units of the 
trap energy, $E_{tur} = 7.4 E_\perp$. In the regime of vortex turbulence, the 
energy of a vortex is yet lower than the energy of vortex interactions.

\subsection{Grain turbulence}

When the number of vortices in the random tangle grows to the critical number 
$N_c^*$, the vortex interactions become very important. Then vortices start 
strongly interacting and destroying each other. This happens when the 
interaction energy between two vortices, at a typical distance $\delta$, 
reaches the energy of a vortex \cite{Tsubota_31,Nemirovskii_32,Pethick_34}. 
This critical number of vortices can be estimated as
\be
\label{15}
N_c^* = \left ( \frac{r_\perp}{\dlt} \right )^2 \;   .
\ee
In the present setup, the energy of a vortex becomes equal to the interaction 
energy of two vortices at the distance $\delta \sim 0.71 \times 10^{-4}$ cm,
which yields $N_c^* \approx 30$. This estimate is again in good agreement 
with numerical simulations and experiment. 

The grains have the linear sizes between $10^{-5}$ cm and $5 \times 10^{-5}$ cm,
which defines the typical size of the grain as $3 \times 10^{-5}$ cm. This 
typical size is of order of the coherence length $\xi$, as it should be for 
a coherent droplet. Their density is much larger than that of their surrounding. 
They look like the droplets of water flowing inside fog. 

The corresponding injected energy, needed for producing grain turbulence, can 
be estimated as $E_{fog} = N_c^* E_{vor}$, which equals 
$E_{fog} = 0.771 \times 10^{-11}$ eV, that is, $E_{fog} = 8.87 E_\perp$.  

Figure 2 presents the sequence of experimentally realized states, observed 
in time-of-flight experiments with $^{87}$Rb: a state with several vortices, 
vortex turbulent state, and the density configuration, typical of a granular 
state. Brighter colour corresponds to higher density. 

Numerical simulations confirm that the grains have the sizes close to the 
coherence length $\xi \sim 10^{-5}$ cm and that the phase inside a grain is 
constant, hence the grains are really Bose-condensed coherent droplets. If 
the pumping by the alternating external potential is stopped, each grain 
survives during the lifetime $10^{-2}$ s, which is much longer than the local 
equilibration time of $10^{-4}$ s. That is, the fog made of these Bose-condensed 
grains can be treated as a metastable object, similarly to the fog made of 
water droplets \cite{Yukalov_44}. The density inside a grain is up to $100$ 
times larger than that of their surrounding. It is worth stressing that the 
grains as well as their surrounding are formed by the same atoms. In that 
sense, the system should not be treated as composed of different species 
\cite{Pinto_37}, but rather it could be interpreted as consisting of different 
phases \cite{Yukalov_26}. 

During the regime of grain turbulence, the grains do not essentially change
their sizes, increasing only very little. Since the transition between vortex 
turbulence and grain turbulence is a crossover, in the grain-turbulent state 
there can occur occasional vortices, although they are rare.

\subsection{Wave turbulence}

After the injected energy becomes comparable to the energy $E_c \equiv k_B T_c$,
where $T_c$ is the Bose-Einstein condensation temperature, all condensate 
is destroyed, and the system represents the state of wave turbulence,
or weak turbulence. For $^{87}$Rb atoms, $T_c = 2.76 \times 10^{-7}$ K.
Hence, $E_c = 0.238 \times 10^{-10}$ eV, which gives $E_c = 27.4 E_\perp$. 

Contrary to grains, the waves in the regime of wave turbulence are 
small-amplitude nonuniformities, being just three times denser than their 
surrounding. The typical size of a wave is of order $10^{-4}$ cm, and the 
phase inside it is random, which implies that there are no coherent regions 
that would be related to Bose-Einstein condensate. In the regime of wave 
turbulence, the Bose condensate is getting destroyed.

Because of the high energy required for creating the wave turbulence, this
regime has not yet been achieved in experiments, but has been realized in 
numerical modeling. The whole sequence of nonequilibrium states, found 
numerically, is presented in Fig. 3.   

Recall that the transitions between the dynamic regimes are crossovers, but
not sharp phase transitions. But, as is seen from Fig. 1 for the number of 
vortices as a function of time, the transitions from the vortex state to
vortex turbulence and from vortex turbulence to grain turbulence are rather 
sharp. Contrary to this, the transformation from grain turbulence to wave
turbulence is rather gradual, the grains and waves coexist in a wide temporal 
range. The number of coherent grains diminishes with time, while the number 
of waves increases. The system coherence is being destroyed in a slow pace. 
First, coherence is destroyed at the surface of the atomic cloud. The
snapshots of the phase are shown in Fig. 4. The crossover line, conditionally 
separating the regimes of grain turbulence and wave turbulence, can be 
located at the moment, when the system coherence, associated with grains, 
is destroyed in approximately half of the system, the incoherent region 
being associated with waves.  

\vskip 2mm

During the process of pumping, the kinetic energy of the system, $E_{kin}$, 
increases, as compared to the energy of atomic interactions $E_{int}$. 
This is illustrated in Fig. 5. In the region of weak nonequilibrium, the 
interaction energy is much larger than kinetic energy. The vortex state,
appearing at $t_{vor} = 18$ ms and ending at $t_{tur} = 28$ ms, is 
characterized by the ratio of kinetic to interaction energy between 
$1.8$ and $4.2$. Vortex turbulence, arising at $t_{tur} = 28$ ms and lasting 
till $t_{fog} = 37$ ms, is described by the ratio $E_{kin}/E_{int}$ between 
$4.2$ and $6.7$. Under grain turbulence, existing in the interval from 
$t_{fog} = 37$ ms to $t_c = 120$ ms, the ratio $E_{kin}/E_{int}$ varies from 
$6.7$ to $74$. In the regime of wave turbulence, kinetic energy is much 
higher than interaction energy, so that $E_{kin}/E_{int}$ is of order 
$50$ to $100$.

\section{Discussion}

When a system, being prepared in a strongly nonequilibrium symmetric state, 
equilibrates on a finite time scale to a state with broken symmetry, there 
appear topological defects, such as vortices, protodomains, cells, or 
grains, strings, and monopoles. This is called the Kibble-Zurek mechanism
that can be illustrated by various systems, from different kinds of condensed 
matter to cosmological objects 
\cite{Kibble_1,Kibble_2,Zurek_3,Zurek_4,Zurek_5,Polkovnikov_38,Yukalov_39}.
In the particular case of nonequilibrium Bose condensation, the equilibration 
process starts with the nonequilibrium state of wave turbulence and ends in 
an equilibrium Bose-condensed state \cite{Zakharov_45,Nazarenko_40}.
 
We have shown that there exists the {\it inverse Kibble-Zurek scenario}, 
when we start with an equilibrium system with broken symmetry and, by 
imposing perturbations, transform it to a strongly nonequilibrium symmetric 
state through the sequence of states with spontaneously arising topological 
defects. This is demonstrated by considering a trapped Bose-Einstein condensate, 
perturbing which we realize a sequence of states whose order is opposite to 
that arising in the process of equilibration of an uncondensed system to 
its Bose condensed state. All intermediate states with topological defects 
are recovered, although in the opposite to the Kibble-Zurek mechanism order. 
Starting with an equilibrium Bose condensate, by increasing perturbation, 
we consecutively observe the creation of separate vortices, vortex turbulence, 
condensed cells, or grains, and wave turbulence with Bose condensate being 
destroyed. These are the same states that would appear in the opposite order, 
if an uncondensed nonequilibrium system would equilibrate to its Bose condensed 
equilibrium. We demonstrate the inverse Kibble-Zurek scenario both 
experimentally, by perturbing the Bose-Einstein condensate of trapped 
$^{87}$Rb atoms, and also by accomplishing numerical simulations for the 
same setup as in the experiment, the experimental and numerical results being 
in good agreement with each other. 
 
The summary of the realized states, depending on the amount of the injected 
energy, is as follows:
\begin{center}
\begin{tabular}{cl}
$E_{inj} = 0 $                 ~~&~~ (equilibrium state) \\
$0< E_{inj} < 2E_{vor}$        ~~&~~ (weak nonequilibrium) \\
$2E_{vor} < E_{inj} < E_{tur}$ ~~&~~ (vortex state) \\
$E_{tur} < E_{inj} < E_{fog}$  ~~&~~ (vortex turbulence) \\
$E_{fog} < E_{inj} < E_c$      ~~&~~ (grain turbulence) \\
$E_{inj} > E_c$                ~~&~~ (wave turbulence) .
\end{tabular}
\end{center}

In terms of temporal scale, the observed nonequilibrium regimes occur in 
the intervals
\begin{center}
\begin{tabular}{cl}
$t = 0 $                 ~~&~~ (equilibrium state) \\
$0< t < t_{vor}$        ~~&~~ (weak nonequilibrium) \\
$t_{vor} < t < t_{tur}$ ~~&~~ (vortex state) \\
$t_{tur} < t < t_{fog}$  ~~&~~ (vortex turbulence) \\
$t_{fog} < t < t_c$      ~~&~~ (grain turbulence) \\
$t > t_c$                ~~&~~ (wave turbulence) ,
\end{tabular}
\end{center}
where $t_{vor} = 18$ ms, $t_{tur} = 28$ ms, $t_{fog} = 37$ ms, and 
$t_c = 120$ ms.   

Introducing the relative injected energy
$$
 \ep \equiv \frac{E_{inj}}{E_{vor}} \;  ,
$$
normalized to the vortex energy, the sequence of the states in the inverse
Kibble-Zurek scenario, we have realized, is characterized by the relations
\begin{center}
\begin{tabular}{cl}
$\ep = 0 $      ~~&~~ (equilibrium state) \\
$0< \ep < 2$    ~~&~~ (weak nonequilibrium) \\
$2 < \ep <25$   ~~&~~ (vortex state) \\
$25 < \ep < 30$ ~~&~~ (vortex turbulence) \\
$30 < \ep < 100$ ~~&~~ (grain turbulence) \\
$\ep > 100$      ~~&~~ (wave turbulence) .
\end{tabular}
\end{center}

The injected energy (3), for the used oscillating perturbation potential is 
approximately proportional to time, after it is much longer than the period
of the modulating field, $t \gg 2 \pi/ \omega$, which is of order $10$ ms.
Thus, we have
$$
\frac{E_{fog}}{E_{tur}} \sim  \frac{t_{fog}}{t_{tur}} \sim 1 \; , \qquad
\frac{E_c}{E_{fog}} \sim  \frac{t_c}{t_{fog}} \sim 3 \; . 
$$

The different dynamic regimes are not separated by sharp boundaries, but, 
strictly speaking, occur continuously. That is, these dynamic transitions 
are crossovers, but not abrupt phase transitions, although they may happen 
rather noticeably. Therefore the neighboring states can occasionally display 
the features typical of their lower-energy member. For instance, in the 
regime of grain turbulence, there may happen occasional vortices. And in 
the regime of wave turbulence, waves coexist with grains.    

Thus we have demonstrated that the Kibble-Zurek mechanism is reversible, at 
least for the process of Bose-Einstein condensation versus Bose condensate 
destruction. The demonstration of the inverse Kibble-Zurek scenario for 
trapped atoms opens a new direction of studying the universality of the 
reversibility of this mechanism, investigating whether it is reversible for 
other physical systems of different nature.   

\vskip 5mm

{\bf Acknowledgments}
\vskip 2mm
We appreciate financial support from FAPESP and RFBR (Grant 14-02-00723). 
We are grateful to M. Tsubota and E.P. Yukalova for discussions.

\newpage

\begin{center}
{\bf{\Large
Figure Captions }}
\end{center}

\vskip 2cm
{\bf Figure 1}. Number of vortices as a function of the pumping time, under 
fixed modulation amplitude, found in numerical simulations.  

\vskip 2cm
{\bf Figure 2}. The sequence of nonequilibrium states observed in experiment: 
(a) vortex state, (b) vortex turbulence, (c) grain turbulence. The density 
snapshots, obtained in the time-of-flight setup, are presented. Darker spots 
correspond to lower density. 

\vskip 2cm
{\bf Figure 3}. The sequence of nonequilibrium states realized in numerical 
modeling: (a) vortex state, (b) vortex turbulence, (c) grain turbulence, 
(d) wave turbulence. The density cross-sections are demonstrated. Brighter 
colour corresponds to higher density. 

\vskip 2cm
{\bf Figure 4}. The sequence of snapshots of the system phase at consecutive
moments of time, the phase being gradually destroyed under increasing 
injected energy. 
 
\vskip 2cm
{\bf Figure 5}. The ratios of interaction energy to kinetic energy of the 
system, $E_{int}/E_{kin}$ (thick line), and of kinetic to interaction energy, 
$E_{kin}/E_{int}$ (thin line), as functions of the time of pumping.

\newpage

\begin{figure}[ht]
\centering
\includegraphics[width=10cm]{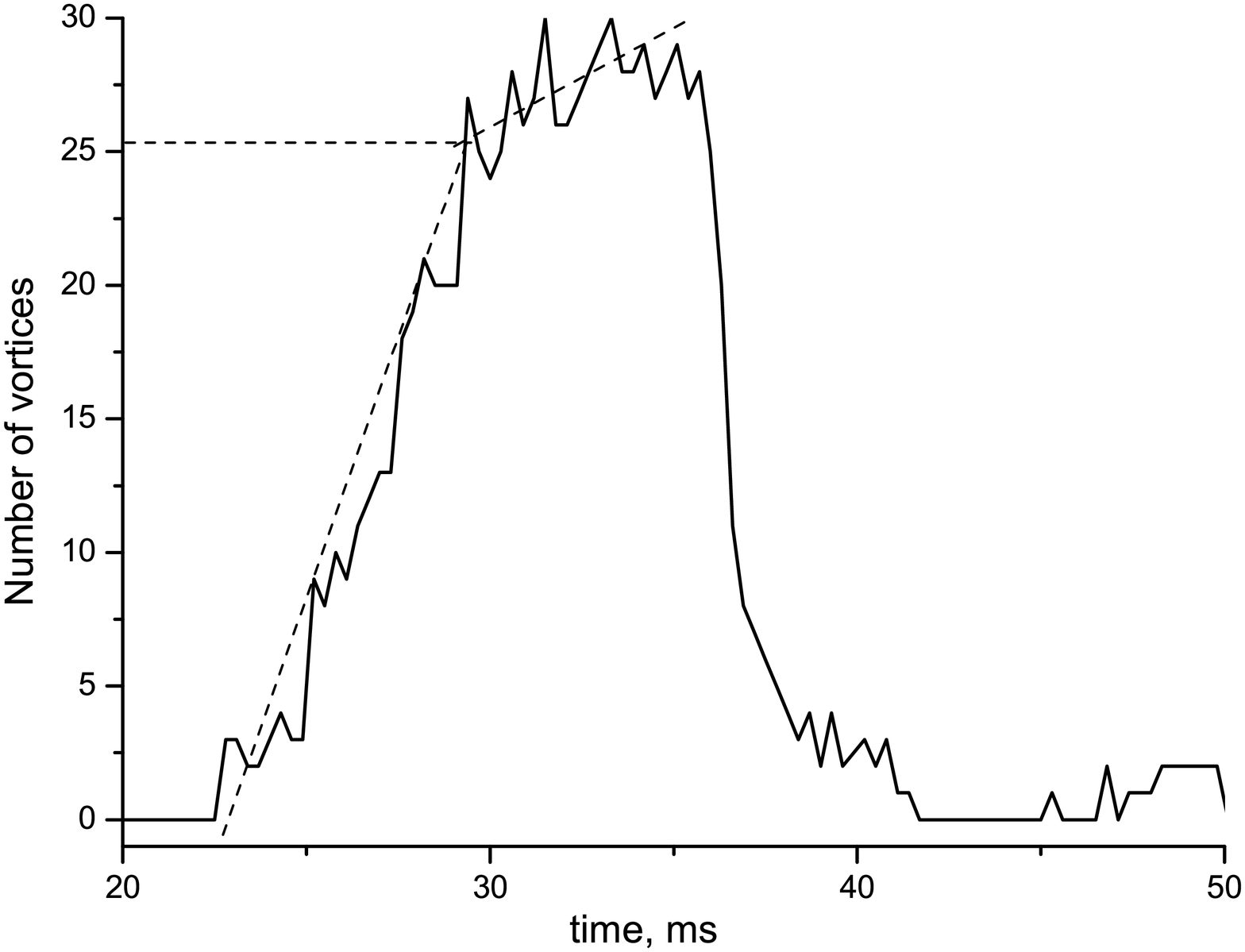}
\caption{Number of vortices as a function of the pumping time, under fixed 
modulation amplitude, found in numerical simulations.}
\label{Fig1}
\end{figure}

\vskip 4cm

\begin{figure}[ht]
\centering
\includegraphics[width=14cm]{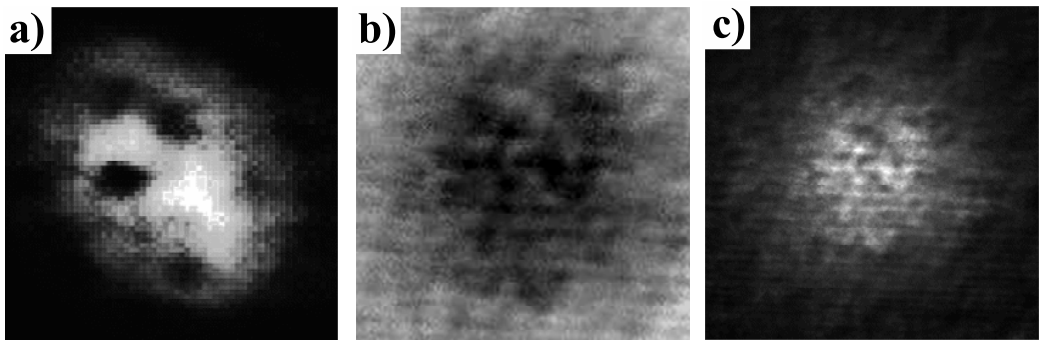}
\caption{The sequence of nonequilibrium states observed in experiment: 
(a) vortex state, (b) vortex turbulence, (c) grain turbulence. The density 
snapshots, obtained in the time-of-flight setup, are presented. Darker spots 
correspond to lower density.}
\label{Fig2}
\end{figure}

\begin{figure}[ht]
\centering
\includegraphics[width=10cm]{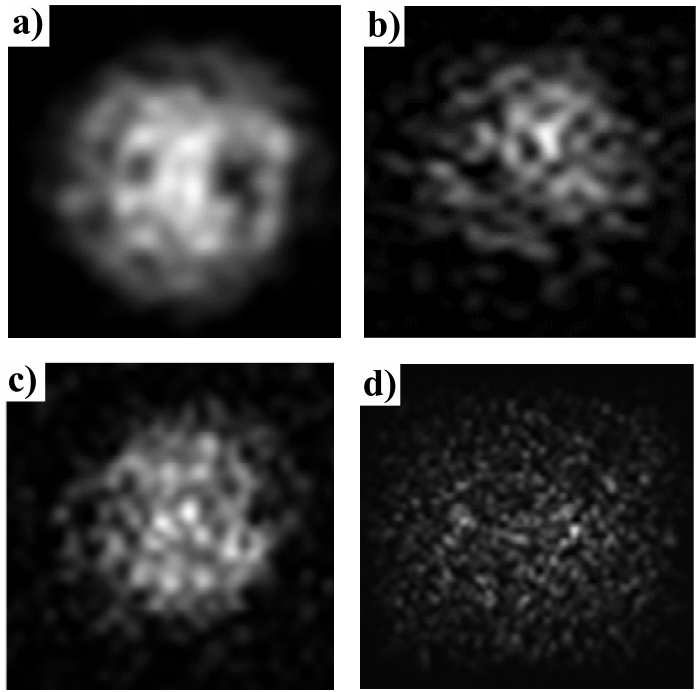}
\caption{The sequence of nonequilibrium states realized in numerical 
modeling: (a) vortex state, (b) vortex turbulence, (c) grain turbulence, 
(d) wave turbulence. The density cross-sections are demonstrated. Brighter 
colour corresponds to higher density.}
\label{Fig3}
\end{figure}

\begin{figure}[ht]
\centering
\includegraphics[width=16.5cm]{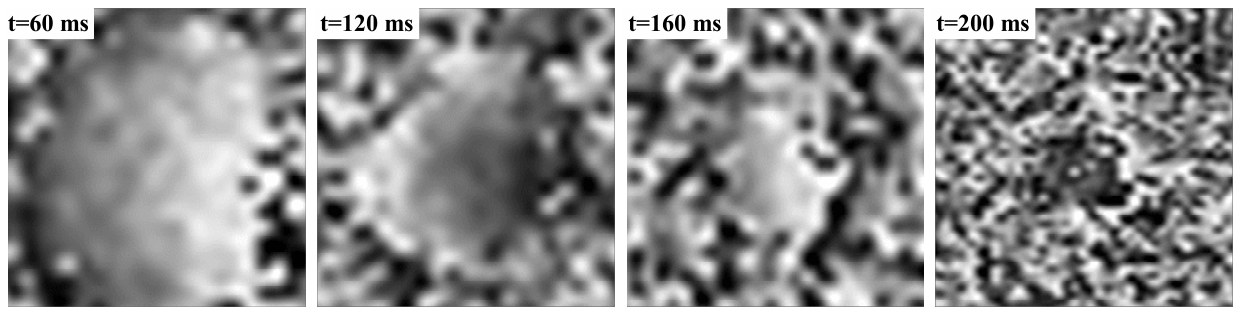}
\caption{The sequence of snapshots of the system phase at consecutive
moments of time, the phase being gradually destroyed under increasing 
injected energy.}
\label{Fig4}
\end{figure}

\begin{figure}[ht]
\centering
\includegraphics[width=10cm]{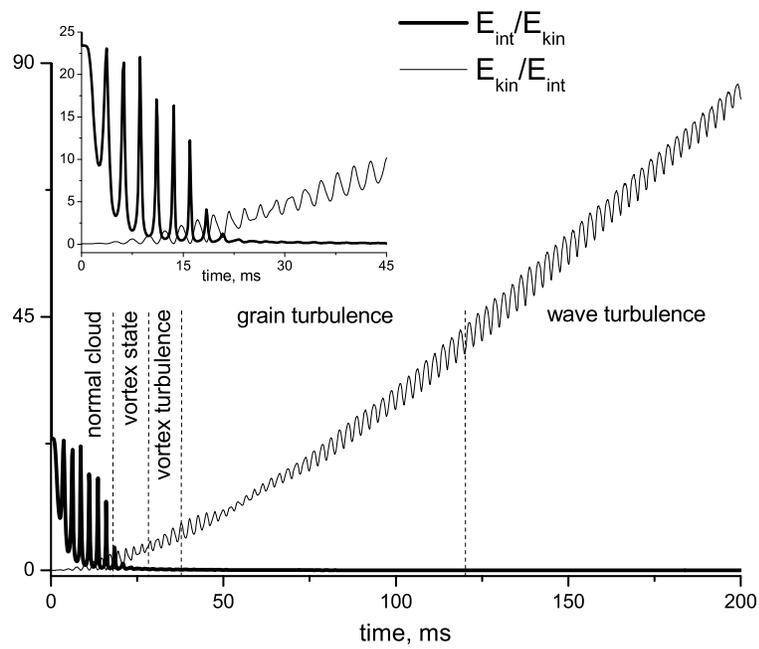}
\caption{The ratios of interaction energy to kinetic energy of the 
system, $E_{int}/E_{kin}$ (thick line), and of kinetic to interaction energy, 
$E_{kin}/E_{int}$ (thin line), as functions of the time of pumping.}
\label{Fig5}
\end{figure}

\end{document}